\documentclass[3p,times,twocolumn]{elsarticle}
\usepackage{ecrc}

\volume{00}
\firstpage{1}
\journalname{Nuclear Physics B Proceedings Supplement}
\runauth{V. Berezinsky}
\jid{nuphbp}
\jnltitlelogo{Nuclear Physics B Proceedings Supplement}
\usepackage{amssymb}

\newcommand{\beq}{\begin{equation}}
\newcommand{\eeq}{\end{equation}}
\newcommand{\ba}{\begin{eqnarray}}
\newcommand{\ea}{\end{eqnarray}}

\newcommand{\AmS}{{\protect\the\textfont2
  A\kern-.1667em\lower.5ex\hbox{M}\kern-.125emS}}
\def\lsim{\raise0.3ex\hbox{$\;<$\kern-0.75em\raise-1.1ex\hbox{$\sim\;$}}}
\def\gsim{\raise0.3ex\hbox{$\;>$\kern-0.75em\raise-1.1ex\hbox{$\sim\;$}}}

\def\theta{\vartheta}

\begin{document}

\begin{frontmatter}

\dochead{}
\title{High Energy Neutrino Astronomy}


\author{V. Berezinsky}

\address{INFN - Laboratori Nazionali del Gran Sasso,
        I--67010 Assergi (AQ), Italy}

\begin{abstract}
The short review of theoretical aspects of ultra high energy (UHE) neutrinos.
The accelerator sources, such as Supernovae remnants, Gamma Ray Bursts, 
AGN etc are discussed. The top-down sources include Topological Defects (TDs), 
Superheavy Dark Matter (SHDM) and Mirror Matter. The diffuse fluxes are 
considered accordingly as that of cosmogenic and top-down neutrinos. Much 
attention is given to the cascade upper limit to the diffuse neutrino 
fluxes in the light of Fermi-LAT data on diffuse high energy gamma radiation. 
This is most general and rigorous upper limit, valid for both cosmogenic and 
top-down models. At present upper limits from many detectors are close to 
the cascade upper limit, and 5~yr IceCube upper limit will be well below it. 
\end{abstract}

\end{frontmatter}

\section{Introduction}
\label{sec:introduction}

Many neutrino telescopes, using different technique of observation, were 
searching for High Energy (HE) cosmic neutrinos during last 30-40 years.
Among existing neutrino telescopes there are deep
underwater/ice detectors (Baikal, ANTARES, IceCube, NESTOR), Extensive 
Air Shower detectors (Auger and HiRes) and radio-telescopes:
ANITA-lite, RICE, GLUE, FORTE and others. Even  bigger projects
include JEM-EUSO and Super-EUSO.  

Why is there such great interest to HE neutrinos? 

HE neutrinos can provide us with most important information in physics
and astrophysics. 

Detection of HE neutrinos from SN remnants will prove that these
objects are sources of galactic cosmic rays (CR) and the Standard
Model of Galactic Cosmic Ray origin will be confirmed. 

Jet models of Gamma Ray Bursts (GRBs) and Active Galactic Nuclei 
(AGN) can be proved. 

Detection of cosmogenic neutrinos can clarify the origin of Ultra High
Energy Cosmic Rays (UHECR)  and determine the model of transition from
galactic to extragalactic CRs. 

Detection of neutrino jets with energies above $10^{20} - 10^{21}$~eV 
means discovery of Topological Defects, important objects in standard
cosmology. 

Registration of HE neutrinos from the center of the Sun or Earth  
indicate the annihilation of Dark Matter (DM) particles there. 

Mirror matter can be discovered with help of oscillation mirror
neutrinos into visible ones. 
\section{Basics of HE neutrino astrophysics}
\label{sec:basics}
We summarize here some basic features of HE neutrino astrophysics.
\\*[2mm]
{\em Production of UHE cosmic neutrinos}\\ 
occurs in $pp$ and $p\gamma$
collisions of UHE protons with the target nuclei and with  
low-energy photons. They can be also produced by annihilation 
of DM particles and by decays of superheavy particles. In all
these cases neutrinos are produced  in the chain of  
pion decays.\\*[2mm] 
 {\em UHE neutrino sources}\\ 
are subdivided into accelerator 
and top-down sources, where neutrinos are produced in decays 
and annihilation of heavy particles. The examples of such sources are 
given by annihilation of neutralinos in the Sun and Earth \cite{Press:1985ug}, 
by topological defects (for a review see \cite{Bhattacharjee:1998qc}), 
which produce superheavy unstable particles, and by decays of quasi-stable 
superheavy DM 
particles \cite{Aloisio:2003xj}. 

Usually HE neutrinos are accompanied by other radiations,
most notably by HE gamma-rays and cosmic rays (CR). There are, however,
so called ``hidden sources'' where all accompanying radiations are 
strongly or fully absorbed. The examples of such objects are the Sun
and Earth, in center of which neutralinos annihilate. Another ideal example  
is given by mirror matter, where all mirror particles interact with 
visible matter gravitationally, and only mirror neutrinos can
oscillate into visible ones. The almost ``hidden''
source is given by the Stecker model \cite{Stecker:1991vm} of AGN, where UHE 
photons and protons are mostly absorbed or confined, and only HE 
neutrinos emerge from there. 

In \cite{Berezinsky:2000bq} a hidden neutrino source is produced in evolution of
stellar cluster at its contraction. Collisions of neutron stars at 
the center of a cluster produce a rarefied cavity filled by
ultra-relativistic fireballs from colliding neutron stars. The cavity
is surrounded by thick gas envelope produced by destructed stars.
All HE particles, such as protons, gamma and X-rays are  
absorbed in the thick envelope, and only UHE neutrinos escape.  

Recently, the ultra-relativistic jet surrounded by the envelope of
collapsing star is considered \cite{Razzaque:2009kq} as hidden UHE neutrino   
source. 

There are also some other more conventional examples of astrophysical 
hidden sources \cite{Berezinsky:1985xp,Ginzburg:1990sk}.\\*[2mm] 
\noindent
{\em Neutrino detection}\\ 
includes four remarkable reactions:\\*[1mm]
\noindent
Muon production $\nu_{\mu}+N \to \mu+ {\rm all}$ gives an excellent tool
to search for the discrete sources, since directions of UHE muon and 
neutrino coincide. \\*[1mm]
\noindent
Resonant production of W-boson, 
$\bar{\nu}_e + e \to W^-\to \bar{\nu}_\mu +\mu$ (the Glashow resonance 
\cite{Glashow:1960zz}) and  $\bar{\nu}_e + e \to W^-\to {\rm hadrons}$ 
\cite{Berezinsky:1977sf} have the large cross-sections. For the practical
applications the latter reaction is more important since it  
results in production of monoenergetic showers with energy 
$E_0=m_W^2/2m_e=6.3\times 10^6$~GeV \cite{Berezinsky:1977sf} and can be
observable in IceCube and future 1~km$^3$ underwater detectors.  
\\*[1mm]
\noindent
Tau production in a detector, $\nu_{\tau}+N \to \tau + {\rm hadrons}$,
is characterized by time sequence of three signals \cite{Learned:1994wg}: 
a shower from prompt
hadrons, the Cherenkov light from $\tau$ and hadron shower from $\tau$-decay.
UHE $\nu_{\tau}$ are absorbed less in the Earth due to
regeneration: absorbed $\nu_{\tau}$ is converted into $\tau$, which decays
producing $\nu_{\tau}$ again. \\*[1mm]
\noindent
Another remarkable phenomenon produced by UHE $\tau$-neutrinos is 
Earth-skimming effect \cite{Fargion:2000iz}, due to which
the Auger observatory obtained the upper limit 
on UHE $\tau$- neutrino flux \cite{Abraham:2007rj}.  

Z-bursts provide a signal from the space, caused by the resonant $Z^0$ 
production by UHE neutrino on DM neutrino \cite{Weiler:1997sh},       
$\nu+ \bar{\nu}_{\rm DM} \to Z^0 \to {\rm hadrons}$.
The energy of the detected neutrino must be tremendous: 
$E_0=m_Z^2/2m_\nu \sim 10^{24}$~eV.\\*[1mm]
\noindent
{\em Neutrino oscillations}\\ 
play the essential role. 
The neutrino flavors $\bar{\nu}_e$ and $\nu_{\tau}$ are inefficiently
produced in the accelerator sources. The flavor
oscillation $\bar{\nu}_\mu \leftrightarrow \bar{\nu}_e$ and 
$\nu_\mu \leftrightarrow \nu_\tau$ can equalize the fluxes of these 
neutrinos. The oscillation length $L(E)$ is given by  
$$
L(E)=\frac{4\pi E}{\Delta m^2}=8.0\left ( \frac{E}{10^{10}~{\rm GeV}}\right )
\left (\frac{10^{-4}{\rm eV}^2}{\Delta m^2}\right )~{\rm pc}
$$
where $\Delta m^2$ are $2.4\times 10^{-3}$~eV$^2$
and $7.7 \times 10^{-5}$~eV$^2$ for atmospheric and solar neutrino 
oscillations, respectively. Thus oscillation length is very short for cosmic
distances and neutrino oscillations are very efficient.
If neutrino flux is produced by decays of pions and muons  with ratio
$\pi^+/\pi^- =1$, the initial neutrino flavor ratio is  
$\nu_e:\nu_{\mu}:\nu_{\tau}=1:2:0$, and  the observed flavor ratio 
(after oscillation)  is $\nu_e:\nu_{\mu}:\nu_{\tau}=1:1:1$ (equipartition).
This is the case of neutrinos produced in $pp$-collisions. In case of 
$p\gamma$ pion production $\bar{\nu}_e$ can be strongly suppressed,
but flavor equipartition after oscillation is approximately
holds. Many cases of flavor ratios with different conditions
(e.g. neutron decay and survival, muon survival etc are studied in 
\cite{Pakvasa:2007dc}. 

The matter neutrino oscillations in the sources and in the Earth 
can also occur (see e.g. \cite{Razzaque:2009kq}).\\*[2mm]
{\em HE neutrinos from early universe}.\\
One might think (and many did think) that large neutrino fluxes can 
be produced at cosmological epochs with large red shift $z$, e.g. due 
to decay of superheavy particles and production by topological defects. 
In fact, this possibility is disfavored \cite{Berezinsky:1991aa,Protheroe:1994dt} 
by absorption of HE 
neutrinos and by nucleosynthesis bound on their fluxes. Neutrinos are 
absorbed in $\nu\bar{\nu}$ collisions with big-bang neutrinos and horizon of 
observation for neutrinos with energy $E_{\nu0}$ (at present) is given
by redshift
$$
z_{\rm abs}=7.9\times 10^4 (E_{\nu 0}/{\rm 1~TeV})^{-1/3}.
$$
Neutrino fluxes produced at large z (e.g. by topological defects) are 
strongly restricted by production of $D$ and $^3He$ at the epochs 
after Big Bang nucleosynthesis. Neutrinos cause e-m cascades and 
MeV photons from these cascades produce $D$ and $^3He$ in
collisions with $^4He$ nuclei.    
\section{Astrophysical (accelerator) sources}
Protons are assumed to be accelerated mostly by the shocks and produce 
neutrinos in $pp$ and $p\gamma$ collisions.\\
{\em HE neutrinos from SN remnants}\\*[1mm]
Detection of HE neutrinos from SN remnants is one of the most important
tasks of HE neutrino astronomy, and this task looks perfectly
realistic for IceCube and other 1~km$^3$ detectors. 
\noindent
Acceleration of protons and nuclei in SN remnants (SNRs) is the basic element
of the Standard Model for Galactic CR (see e.g. \cite{2007ApJ...661L.175B}). 
These sources successfully explain the spectra and fluxes of CR
observed in our Galaxy, and the knee is interpreted 
as the end of Galactic CR.  Accelerated protons
interacting with the gas in a SNR  must emit gamma-rays and 
neutrinos through production and decays of neutral and charged pions. 
HE gamma rays are detected from several SNRs, but from most of    
them the signal is compatible with the bremsstrahlung or Inverse 
Compton production by  HE electron. There are in some cases the 
indications to the {\em hadronic} gamma-rays 
(i.e. produced by neutral pions born in $pp$-collisions), however 
in all these cases one still may argue in HE electron production of the
observed signal. Detection of HE neutrino signal gives unambiguous prove 
of pion production by accelerated protons. 

There are at least two SN remnants, W28 and W44, for which the
evidence for hadronic gamma-ray signal is rather strong. 
For both of them gamma-radiation is produced in a nearby dense molecular 
cloud. Gamma radiation from W28 was 
observed by AGILE detector at $E > 400$~MeV 
\cite{Giuliani:2010fp}, and by Fermi-LAT \cite{Abdo:2010vj} and 
H.E.S.S \cite{HESS} at higher energies. The proof is based on
spectral characteristics and  molecular cloud as the target. 
The case of W44 observed by Fermi-LAT is similar: gamma-radiation is
observed from dense cloud ring around  W44 \cite{Abdo:2009zz}.   

SNRs discussed above are candidates for detectable neutrino fluxes by 
IceCube. Another candidate for the detected hadronic gamma-rays 
SNR RX J1713.7-3946 was analyzed in \cite{Vissani:2009vv} for detectable
neutrino flux. The authors accurately recalculated the observed
gamma-ray flux, assuming its hadronic origin, to HE neutrino flux,
and found that it corresponds to 2 - 3 events per year at energy 
above 1~TeV for a detector as IceCube. 
{\em Gamma Ray Bursts}.\\*[1mm]
\noindent
GRBs are most exiting sources of UHE neutrinos. There are two
mechanisms of HE neutrino generation. In the first one \cite{Vietri:1998nm} 
particles are accelerated by external shock and neutrinos are produced 
in $p\gamma$ collisions with GRB photons behind the shock. In the
second mechanism \cite{Waxman:1997ti} protons are accelerated by internal 
shocks, with the spectrum assumed to be $\propto 1/E^2$.  Neutrinos 
are produced in $p\gamma$ collisions with GRB photons. All estimates are 
very transparent and follow from assumption that the energy outputs 
in GRB photons, accelerated protons and produced neutrinos are about
the same: $W_{\nu} \sim W_p \sim W_{\rm GRB}$. Then the total number
of neutrinos with energy $E$ per burst is 
$$
N_{\nu}(E) \sim \frac{W_{\rm GRB}}{\ln E_{\rm max}/E_{\rm min}}E^{-2},
$$
where $E_{\rm min}$ and $E_{\rm max}$ are minimum and maximum acceleration
energy, respectively. Now one can express the flux of neutrinos from a
single GRB, in terms of neutrino fluence $S_{\nu}$:
$$
F_{\nu_{\mu}+\bar{\nu}_{\mu}}(E)=\frac{1}{3}~ 
\frac{S_{\nu}}{\ln\;(E_{\rm max}/E_{\rm min})}\;E^{-2},
$$
and calculate the number of muons produced in 1~km$^3$ detector 
per burst as 
$$
P_\mu = N_n \int dE\; F_{\nu_{\mu}+\bar{\nu}_{\mu}}(E)\; \sigma_{\nu N} (E),
$$
where $N_n=6\times 10^{38}$ is the number of nucleons in the detector 
and $\sigma_{\nu N}$ is $\nu_{\mu} N$ - cross-section. 
For $S_\nu \sim 10^{-5}$~erg/cm$^2$, with frequency of bursts as in
Fermi GRB Monitor $\dot{N}_b \sim 500$~yr$^{-1}$, one finds the number
of produced muons with $E \geq 10$~TeV  for 5~yr as 
$P_\mu \dot{N}_b t \sim 0.06$, i.e. too low for detection by IceCube.

The diffuse flux is estimated in identical way through the local neutrino 
emissivity ${\cal L}_{\nu}(0)$ and evolutionary factor $k_{\rm evol}$ 
$$
J_{\nu_\mu+\bar{\nu}_{\mu}}(E)=\frac{1}{3}\frac{cH_0^{-1}}{4\pi}
\frac{{\cal L}_{\nu}(0)}{E^2\ln E_{\rm max}/E_{\rm min}} k_{\rm evol},
$$
where evolutionary factor is given by 
$$
k_{\rm evol}= \int_0^{z_{\rm max}}\frac{dz}{(1+z)^2}\frac{f_{\rm evol}(z)}
{\sqrt{(1+z)^3\Omega_m+\Lambda}}.
$$
For the evolutionary function $f_{\rm evol}(z)$ we take the case
of strong star-formation evolution from \cite{Wick:2003ex}, which results in 
$k_{\rm evol}=7.0$ as a maximum value. For the emissivity we use   
\vspace{-2mm}
$$
{\cal L}_{\nu}(0) \leq {\cal L}_{\rm GRB}(0)= 0.6\times 10^{43}~
{\rm erg}/{\rm Mpc}^3{\rm yr},\\*[-2mm]
$$
\noindent
where the local GRB emissivity is taken from \cite{schmidt}. 
The most recent estimate 
$0.5\times 10^{43}~{\rm erg}/{\rm Mpc}^3{\rm yr}$ \cite{Eichler:2010ky}
agrees with the value above. For more detailed discussion 
see \cite{Berezinsky:2002nc}.

As a result $E^2J_{\nu_\mu+\bar{\nu}_{\mu}}(E)=
1.15$~eV/cm$^2$ s sr, i.e. 25 times lower than sensitivity 
of IceCube shown in Fig.\ref{fig:upper-limits}.   

GRB neutrinos are detectable by IceCube in case the hadronic energy output  
is an order of magnitude higher than one observed in photons, but it
further strengthens the energetic deficit in GRB models. From observational
point of view the signature of neutrino observation is very reliable
due to time and direction correlation with gamma-radiation. \\*[2mm]
{\em HE neutrinos from AGN jets}.\\*[1mm]
\noindent
The models of HE neutrino production in the AGN jets are very similar
to that for GRBs. The protons are accelerated by the multiple shocks 
in the AGN jets, especially in their inner parts. Neutrinos are produced 
in the collisions with
photons from the accretion disc and from photons produced in the jet 
by accelerated electrons and protons. Neutrino flux from individual 
AGN is very small, but the diffuse flux is predicted to be detectable 
by IceCube detector. The estimates can be performed similar to those 
in subsection above. One can found the detailed calculations in 
\cite{Atoyan:2002gu,Dermer:2003rt,Dermer:2008cy}. \\*[2mm]   
{\em HE neutrinos from galaxy clusters} \cite{cluster}.\\*[1mm]
The clusters of galaxies are able to confine the UHE particles for a time 
exceeding the age of the universe. This is the key phenomenon which 
makes galaxy clusters the powerful sources of UHE neutrinos. The
particles are accelerated in clusters by various mechanisms: 
in the normal galaxies by SN shocks, in AGN and cD-galaxies, in 
the process of galactic merging etc. 
The diffuse HE neutrino flux is determined
entirely by basic parameters characterizing the clusters. In particular,
for the lower limit of the diffuse flux provided  by normal galaxies
in a cluster with CR luminosity $L_p$ and generation index $\gamma_g$, 
both taken as ones in our galaxy, the diffuse flux is given as 

$$
J_{\nu}(E) \propto L_p E^{-\gamma_g}\frac{N_g}{R_{\rm cl}^3}\xi 
\Omega_b\rho_{\rm cr},
$$

\begin{figure}[t]
\begin{center}
\includegraphics[width=8cm,height=6cm]{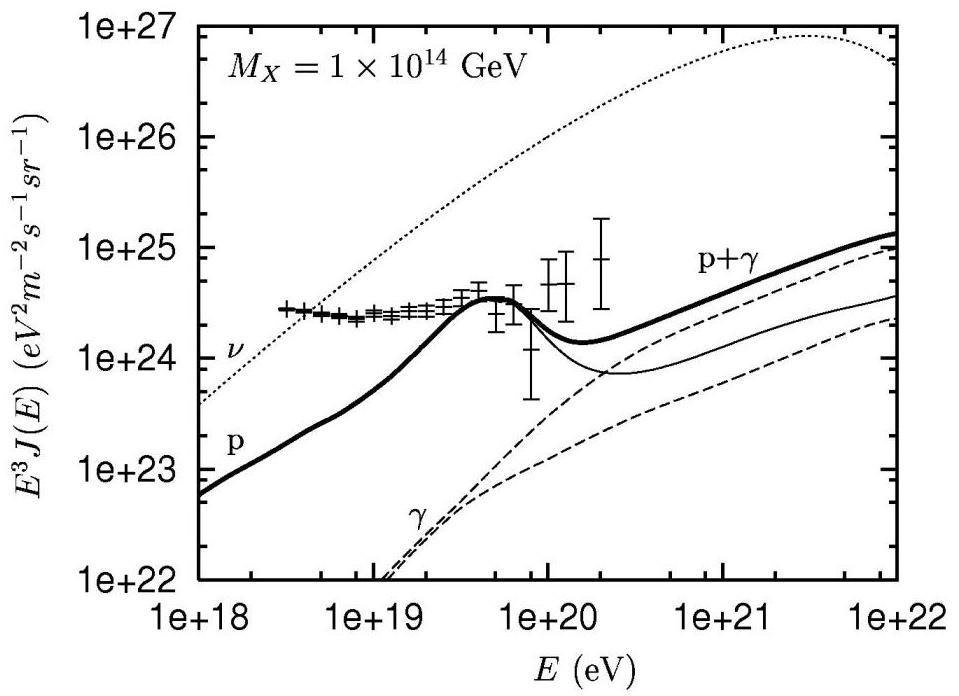}
\end{center}
\vskip -5mm
\caption{Diffuse all-flavor neutrino spectrum from necklaces for 
$m_X=1\times 10^{14}$~GeV \cite{Aloisio:2003xj}. The thick curve gives $p+\gamma$ 
flux normalized to the
AGASA UHECR data. If to normalize the proton flux in this figure 
by HiRes data all curves, including one labeled $\nu$ (neutrino flux),
should be lowered by factor 3 - 5. }
\label{neckl}
\end{figure}
\noindent
where $R_{\rm cl} \sim 2$~Mpc is the virial radius of a cluster, 
$N_g \sim 100$ is richness of a cluster, $\rho_{\rm cr}$ is critical
cosmological density, and $\xi\Omega_b$ is cosmological baryonic density
provided by clusters. The flux is marginally detectable by IceCube. \\*[2mm]
\section{Non-accelerator neutrino sources} \label{non-acc}
These sources include objects with annihilation of DM (the Sun, Earth, 
cores of the galaxies), objects with the decays of SHDM 
particles (galactic halos) and TDs. In the last two cases  
neutrinos are produced in the decays of superheavy particles with the 
masses up to $M_{\rm GUT} \sim 10^{16}$~GeV. 
\begin{figure}[t]
\begin{center}
\includegraphics[width=8cm,height=6cm]{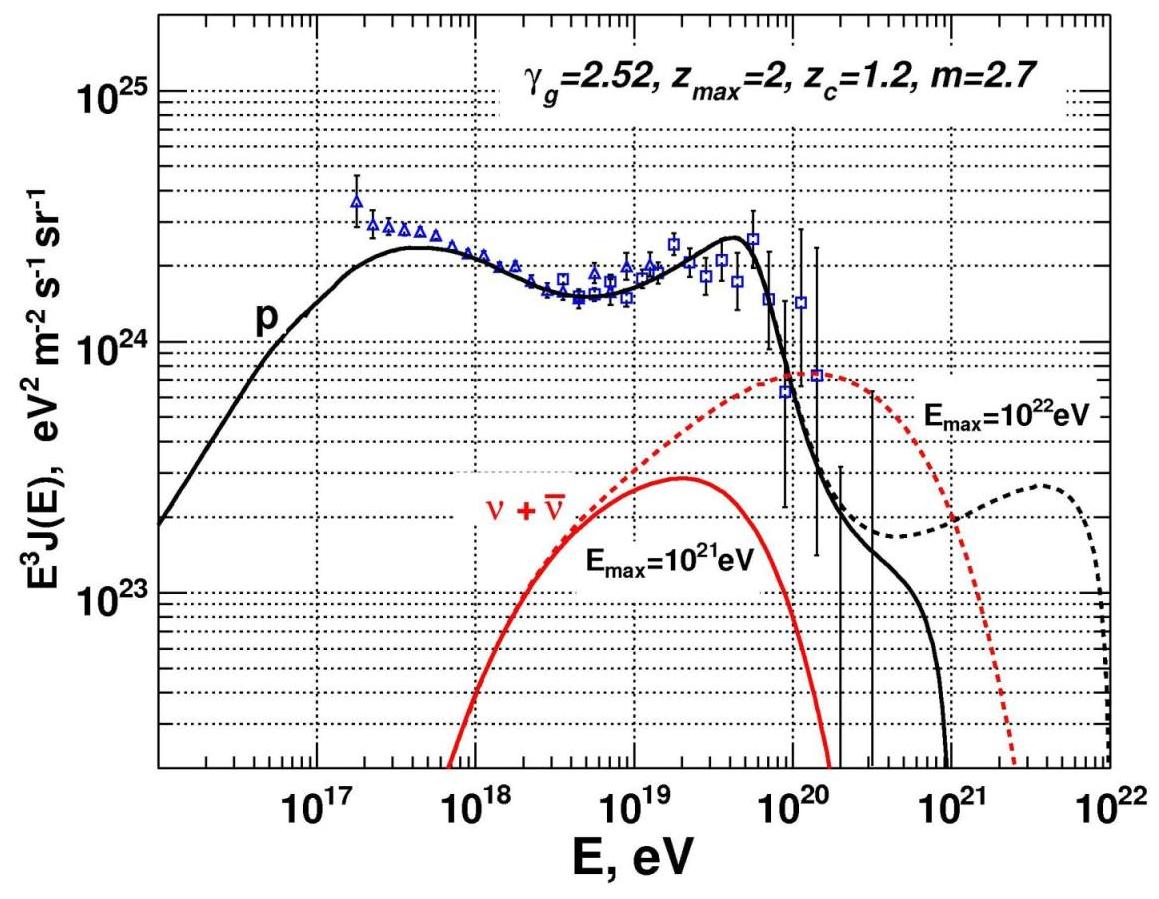}
\end{center}
\vskip -5mm
\caption{UHE neutrino flux in the dip model with AGN as the sources of 
UHECR. The cosmological evolution of AGN with $m=2.7$ up to 
$z_c = 1.2$  is taken from  X-ray observations of AGN. At larger $z$
the evolution is frozen up to $z_{\rm max}=2.0$. The fit of the dip 
is very good, though requires  $\gamma_g = 2.52$ different from the 
non-evolutionary case $m=0$. The neutrino fluxes are given for one
neutrino flavor. 
}
\label{fig:AGN}
\end{figure}
Neutrino spectrum can be approximately described at 
highest energies as $dE/E^{2}$. \\*[2mm]
{\em Neutralino annihilation in the Sun and Earth}.\\*[1mm]
Neutralino is the best motivated DM particle. Crossing the 
Sun or Earth a neutralino can loose its energy in collisions with 
nuclei and diminish its velocity below the escape velocity. If it
happens, a neutralino becomes gravitationally trapped in the object, and  
loosing further their energies, neutralinos are accumulated in the
center of a celestial body \cite{Press:1985ug}. Annihilating there they produce
short-lived hadrons, e.g. D-mesons, which decay to neutrinos.
The process of annihilation strongly depends on 
neutralino mass and composition (mixture of basic fields: zino, bino
and two higgsinos). \\*[2mm]
{\em Superheavy Dark Matter (SHDM)}\\*[1mm]
The first proposal of SHDM \cite{SHDM} was motivated by Ultra High 
Energy Cosmic Rays (UHECR) and by natural character of DM production 
at the epochs soon after inflation. In particular SHDM particles can 
be produced gravitationally \cite{grav-prod}, when the
Hubble parameter $H(t)$ exceeds the particle mass $H(t) \gsim
m_X$. The observed density of DM in the universe 
$\Omega_{\rm cdm} \approx 0.23$, determines the mass of the particle
as $m_X \sim 10^{13}$~GeV. The SHDM particles (X-particles) can be
stable (due to e.g. discrete gauge symmetry) or quasi-stable (due 
to superweak discrete gauge symmetry breaking). The energy spectrum of 
produced particles has approximately power-law form at the highest
energies $\propto E^{-1.9}$ \cite{Aloisio:2003xj}. The dominant decay particles 
are photons
and neutrinos. As any cold DM, X-particles are accumulated in the halos 
of galaxies, in particular in our galaxy with overdensity $2.1\times 10^5$.  
One can expect the detectable fluxes of UHE photons and neutrinos from 
the Galactic Center region.\\*[2mm]  
{\em Topological defects} (TDs). \\*[1mm]
TDs are fundamental cosmological objects. They are produced in early 
universe due to symmetry breaking accompanied by the phase transitions.
In many cases TDs become unstable and decompose to constituent fields,
superheavy gauge and Higgs bosons (X-particles), which then decay
producing UHE neutrinos (see \cite{Bhattacharjee:1998qc,berez99} for the 
reviews).\\*[1mm]
{\em Ordinary strings} are produced by $U(1)$ symmetry breaking.  
There are several mechanisms by which ordinary strings 
can emit HE neutrinos: collapse of the string loops, self-intersection,
annihilation of cusps, production and annihilation of tiny loops.
In most cases produced neutrino fluxes are too low for detection.
More promising scenario is  given by the radiation of
X-particles through the {\em cusp}, a peculiar point where velocity 
reaches velocity of light. This point appears on a loop during each
period. The points of the loop in the cusp region have distribution in 
values of the Lorentz factors $\Gamma$ from a maximum value in the 
cusp point to $\Gamma=1$. The particles escaping through a cusp region 
are boosted by these Lorentz factors. Recently, such case 
has been considered in \cite{vachaspati}. The bosonic (Higgs) condensate 
in a string loop, emits Higgses through the cusp, and due to the
Lorentz factor boost, these particles can reach the tremendous
energies.\\*[2mm]  
{\em Superconducting strings}\\ 
can be powerful sources of neutrinos. 
In a wide class of elementary particle models, strings behave like 
superconducting wires. The charge carriers are massless inside the
string and superheavy outside.  Moving through cosmic magnetic fields, 
such strings develop electric current. When the current reaches the
critical value, the charge carriers escape from a string, turn into 
massive mode and decay. This process is strongly enhanced near the 
cusps due to the Lorentz boost. The decay products, in particular
neutrinos, are emitted isotropically in a frame of cusp segment at rest,   
and propagate in the laboratory system as a very narrow jet with the 
opening angle $\theta \sim 1/\Gamma$.  This scenario was analyzed 
numerically in \cite{olum} with two main model features included. 
First, from all known structures of the universe, the excitation of electric 
current occurs most efficiently in clusters of galaxies for which the magnetic 
field reaches $B \sim 10^{-6}$~G and filling factor $f\sim 10^{-3}$. Second,
the symmetry breaking scale  of order $10^9 - 10^{12}$~GeV must be 
assumed for detectable neutrino fluxes. The typical 
Lorentz factor of the radiating cusp segment is calculated to be 
$\Gamma_c \sim 10^{12}$, and the maximum energy of emitted particle 
can reach $\Gamma_c \eta \sim 10^{22}$~GeV. The neutrino spectrum
is assumed $\propto 1/E^2$, similar to $\propto E^{-1.9}$ \cite{Aloisio:2003xj}. 
The spectrum $E^2J_\nu(E)=$const can be very close to 
the  $E^{-2}$-cascade upper limit in Fig.\ref{fig:upper-limits}. \\*[2mm]
{\em Necklaces (monopoles connected by string)}\\ 
are produced in the $G \to H\times U(1) \to H \times Z_2$ 
sequence of symmetry breaking, with 
each monopole being attached to two strings, and a 
loop reminds a necklace with monopoles playing the role of beads.
In the process of evolution
the strings shrink due to gravitational radiation and
\begin{figure}[t]
  \begin{center}
  \includegraphics[width=8cm]{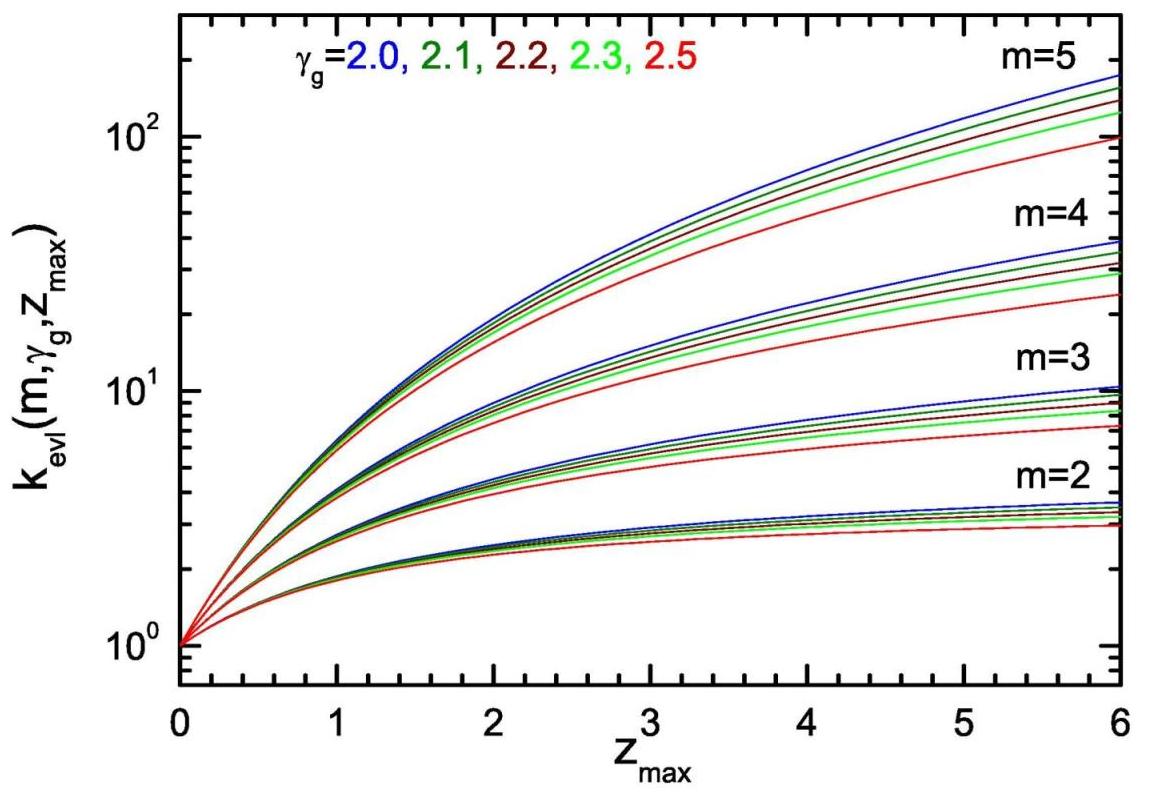}
  \end{center}
  \caption{Evolution factor $k_{\rm ev}$, which describes 
  increasing of neutrino flux due to cosmological evolution 
  of the sources. The evolution factor is shown as function of 
  $z_{\rm max}$ for different indices $\gamma_g$ and parameters 
  of evolution $m$. The evolution factor $k_{\rm ev}$ is large 
  for large $m$ and $z_{\rm max}$.   
  }
    \label{fig:evolution}
  \end{figure}
$M \bar{M}$ pairs in the necklace inevitably annihilate. This model 
is most plausible and well developed for UHE neutrino production. 
Diffuse neutrino flux from necklaces are shown in Fig.~\ref{neckl} 
according to calculations in \cite{Aloisio:2003xj}.

\section{Cosmogenic neutrinos}
Starting from pioneering work \cite{BZ} the fluxes of cosmogenic neutrinos 
have been calculated in many works \cite{ESS} - \cite{olinto-nu}. 
The predicted fluxes differ very considerably, depending on the
assumptions about mass composition of accelerated particles, 
on maximum energy of acceleration and on cosmological 
evolution of the sources. We present here the UHE neutrino
fluxes calculated in the {\em dip model} for the  observed 
UHECR \cite{Berezinsky:2002nc,BGG-PL}, assuming AGN as the sources \cite{BGG-AGN}. 
This model is valid for the  proton-dominated 
composition of UHECR, based on observations of HiRes.  
The pair production dip is a feature of 
interaction of extragalactic UHE protons propagating through CMB.
It is caused by energy losses of
protons due to $p+ \gamma_{\rm CMB} \to e^++e^-+p$ scattering. This feature
in proton spectrum is well confirmed by observational data 
\cite{Berezinsky:2002nc,BGG-PL}.

To calculate neutrino flux produced by UHE protons it is enough 
to know the generation rate of UHE protons at each cosmological epoch. 
We take it as $Q(E)(1+z)^m$, where $Q(E) \propto E^{-\gamma_g}$ and 
$(1+z)^m$ describes the cosmological evolution of the sources
up to some maximal redshift $z_{\max}$. In calculations we consider 
two cases: without evolution when we have only one free parameter, the
generation index $\gamma_g$ , and evolutionary scenario with three free
parameters $\gamma_g$, $m$ and $z_{\max}$. We must fit the observed HiRes 
spectrum with one calculated at 
$z=0$. In non-evolutionary scenario the best fit with very good 
$\chi^2$ is given by $\gamma_g = 2.7$. For the fit with evolutionary model   
we assume AGN as the sources and take the AGN evolution from X-ray
observations \cite{AGN-evolution}: $(1+z)^m$ with $m=2.7$ up to 
$z_c=1.2$, and frozen evolution from $z_c$ to $z_{\rm max}=2$. 
The generation index is fixed as $\gamma_g = 2.52$ for the best fit 
of HiRes data (see Fig.~\ref{fig:AGN}). One may notice that the
theoretical dip automatically describes the ankle at 
$E \approx 5\times 10^{18}$~eV. The calculated one-flavor neutrino fluxes  
for the AGN evolutionary model are shown in Fig.~\ref{fig:AGN} for 
two values of $E_{\max}$.

The evolution is able to increase strongly the neutrino flux.   
The increase of the flux is given by the evolution factor 
$k_{\rm ev}$,  which depends on $m$,~ $z_{\max}$ and $\gamma_g$. 
This dependence is shown in Fig.~\ref{fig:evolution}. 
\section{Cascade upper limit on diffuse neutrino flux}
The e-m cascade upper bound puts the rigorous  upper limit on
UHE neutrino flux \cite{BS,Ginzburg:1990sk}. This limit, in contrast to WB upper
limit \cite{WB}, is valid for both accelerator and non-accelerator 
neutrinos. The production of neutrinos  
is accompanied by production of high energy photons and 
electrons from pion decays. Colliding with low-energy target photons, 
a primary photon or electron produces
e-m cascade due to reactions $\gamma+\gamma_{\rm tar} \to e^++e^-$,
$e+\gamma_{\rm tar} \to e'+\gamma'$, etc. 
The cascade spectrum in its high-energy part is proportional to $E^{-2}$,
which is very close to the EGRET observations in the range 
10~MeV - 100~GeV \cite{EGRET}. The observed energy density in this range is 
$\omega_{\rm EGRET} \approx (2 - 3)\times 10^{-6}$~eV/cm$^3$. The
cascade energy density must be $\omega_{\rm cas} \leq \omega_{\rm EGRET}$,  
and it limits diffuse neutrino flux. 
The situation has dramatically changed with the new data of Fermi-LAT
\cite{fermi-lat} on the flux and spectrum of diffuse 
extragalactic gamma-radiation. In comparison with EGRET this flux is
lower and spectrum is steeper ($\propto E^{-2.4}$). It results in 
stronger upper limit on the cascade energy density 
$\omega_{\rm cas} \leq 5.8 \times 10^{-7}$~eV/cm$^3$ \cite{BGKO},
which severely diminishes the allowed UHE neutrino fluxes 
\cite{BGKO,sarkar}. 
\begin{figure}[t]
\begin{center}
\includegraphics[width=8cm,height=6cm]{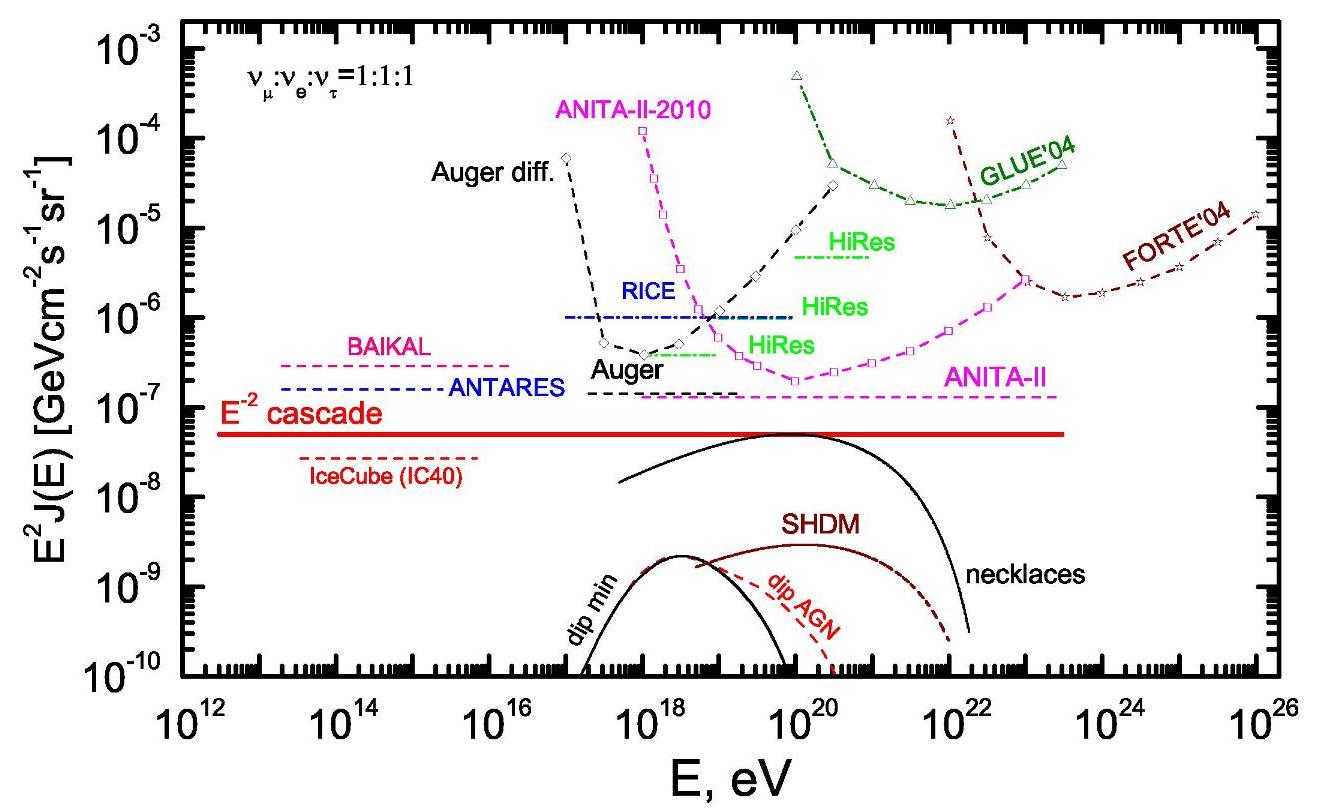}
\end{center}
\vskip -4mm
\caption{ The experimental upper limits on UHE neutrino fluxes 
in comparison with e-m cascade upper limit in assumption of 
$E^{-2}$ generation spectrum (curve $E^{-2}$ cascade) and with predictions 
for cosmogenic neutrinos in the dip model (curves dip-min and dip-AGN), 
for neutrinos from necklaces and from SHDM.  
Neutrino fluxes from necklaces and SHDM are normalized by AGASA data, 
and for normalization by HiRes data the fluxes should be diminished by 
factor 3 - 5. Neutrino flux from superconducting strings is given by 
$E^2 J(E)=$const and it can reach the upper limit '$E^{-2}$cascade'.
Neutrino fluxes are given for three flavors.    
}
\label{fig:upper-limits}
\end{figure}
The maximally  allowed cascade energy density 
$\omega_{\rm cas}^{\rm max} \approx 5.8 \times 10^{-7}$~eV/cm$^3$
provides the upper limit on the integral UHE neutrino flux 
$J_{\nu}(>E)$ (sum of all flavors). It is given by chain of the 
following inequalities  
$$
\omega_{\rm cas}>\frac{4\pi}{c}\int_E^{\infty}EJ_{\nu}(E)dE>
\frac{4\pi}{c}E\int_E^{\infty}J_{\nu}(E)dE \nonumber
\label{eq:int-limit}
$$
where $\omega_{\rm cas} < \omega_{\rm cas}^{\rm max}$,
and the integral in rhs of Eq.~(\ref{eq:int-limit}) 
gives the  integral spectrum of neutrinos $J(>E)$. Thus, 
this equation gives the upper limit on the integral neutrino flux, 
which can be expressed in terms of the upper limit on differential 
neutrino spectrum $J_{\nu}(E)$ as
\begin{equation}
E^2 J_{\nu}(E) < \frac{c}{4\pi}\omega_{\rm cas}^{\max}.
\label{cas-rig}
\end{equation}

Eq.~(\ref{cas-rig}) gives the {\em rigorous} upper limit on the 
neutrino flux. 
It is valid for neutrinos produced by HE protons, by topological 
defects, by annihilation and decays of superheavy particles, 
i.e. in all cases when neutrinos are produced through decay of 
pions and kaons.  It holds for arbitrary neutrino spectrum 
falling down with energy. If one assumes some specific shape 
of neutrino spectrum, the cascade limit becomes stronger.   
For $E^{-2}$ generation spectrum, which is used for
analysis of observational data  one obtains the stronger upper
limit.  Given for three neutrino flavors it reads  
\begin{equation}
E^2J_\nu(E) \leq \frac{c}{4\pi}\frac{\omega_{\rm cas}^{\max}}
{\ln (E_{\rm max}/E_{\rm min})},
\label{cas-E2}
\end{equation}
This upper limit is shown in Fig.~\ref{fig:upper-limits}.

The most interesting energy range in Fig.~\ref{fig:upper-limits} 
corresponds to 
$E_{\nu} > 10^{21}$~eV, where acceleration cannot provide protons 
with sufficient energy for production of these neutrinos.
At present the region of $E_{\nu} > 10^{21}$~eV, and   
$E_{\nu} \gg 10^{21}$~eV is considered as a signature of 
top-down models, which provide these energies quite naturally. 

As one can see from Fig.~\ref{fig:upper-limits} the observational 
upper limit for IceCube after 5~yr of observations will be below 
the cascade upper limit. Crossing it, this detector 
will enter the physically allowed region of neutrino fluxes, and it can be 
regarded as historical event.  
The WB upper limit is not relevant for UHE neutrinos: it is not valid
for top-down scenarios because proton production is strongly
suppressed for top-down sources, and it is very uncertain for 
cosmogenic neutrinos, where for the same proton flux the fluxes of
accompanying neutrinos may differ by one-two orders of magnitudes (see 
Fig.~\ref{fig:AGN}). However, the WB upper bound remains the
convenient low-flux benchmark for detection of neutrino fluxes.\\*[2mm]
{\em Mirror matter and mirror neutrinos.}\\*[1mm]
Mirror neutrinos give the only example of fluxes not limited by the
cascade upper limit. 

The concept of mirror matter, as first was suggested by Lee and 
Yang \cite{Lee-Yang}, consists in existence of sector of matter 
fully symmetric with ordinary one and generated by space-reflection    
transformation. Kobzarev, Okun and Pomeranchuk \cite{KOP} added the
basic assumption that these two sectors communicate only
gravitationally. The gravitational interaction results in mixing of
mirror and ordinary neutrinos and their oscillations \cite{osc}. 
In two-inflaton cosmological model \cite{BV-mirror} the mirror matter 
is suppressed, while mirror TDs can strongly dominate.  Mirror TDs 
copiously produce mirror neutrinos with extremely high
energies, which oscillate into visible ones, while all other 
mirror particles, which accompany production of mirror neutrinos,
remain invisible for our detectors. Therefore, the upper limits on HE  
neutrinos in our world do not exist and their fluxes can be above the 
upper limit shown in Fig.~\ref{fig:upper-limits}. Neutrinos from TDs 
typically have very high energies and one can see that fluxes of
discussed neutrinos are now very severely constrained by ANITA-lite 
data \cite{ANITA-lite}. 
\section{Conclusions}

UHE neutrinos are expected to solve many problems in astrophysics 
and cosmology.  Detection of HE neutrinos from SNRs 
by IceCube is needed for confirmation of Standard Model for GCRs. 
UHE neutrinos from GRBs will clarify the nature of these most unusual 
and controversial objects. As far as HE radiation is concerned, this
is true for AGN, too.  

The diffuse UHE neutrino radiation is presented by cosmogenic 
and top-down neutrinos, in particular neutrinos from TDs.
The fundamental problem of astrophysics involved in cosmogenic  
neutrinos is acceleration of particles. The shock 
acceleration at present knowledge of its theory cannot provide 
$E_{\max}$  higher than $10^{21} - 10^{22}$~eV, and thus
energies of cosmogenic neutrinos cannot exceed $3\times 10^{20}$~eV. 
TDs naturally produce neutrinos emitted from cusps with energies by many
orders of magnitude higher. Detection of neutrinos with these energies 
mean discovery of new physics.  

Cascade upper limit is very general bound valid for both cosmogenic
and top-down neutrinos. This upper limit became stronger with new 
Fermi-LAT data on extragalactic HE diffuse gamma-radiation.
From all existing detectors only IceCube reached the sensitivity below the
cascade upper limit (see Fig.~\ref{fig:upper-limits}) and entered 
the physically allowed region for detectable neutrino fluxes. 
It can be considered as historical event for HE neutrino astronomy.   

The flux of cosmogenic neutrinos can be large only in case UHECR are 
proton-dominated. Even in this case the flux is detectable if maximum 
acceleration energy $E_{\max}$ is large and sources have strong 
cosmological evolution (see \cite{BGKO,sarkar}).

Cosmogenic neutrinos of highest energies 
are detectable by future experiment JEM-EUSO in rather extreme 
models with large $E_{\max}$ and strong cosmological evolution 
(see \cite{BGKO,sarkar}). 

The search for UHE neutrinos in any case is a search for a new 
physics, either for astrophysics (the new acceleration mechanisms 
and cosmological evolution of the sources) 
or for topological defects, mirror topological defects and superheavy
dark matter.

\section{Acknowledgments}
I am grateful to Askhat Gazizov for many discussions and help in preparing 
the figures.

\end{document}